\documentclass[journal]{IEEEtran}

\usepackage{cite}
\usepackage{amsmath,amssymb,amsfonts}
\usepackage{algorithmic}
\usepackage{graphicx}
\usepackage{textcomp}
\usepackage{color}
\usepackage{subfiles}
\usepackage{float}
\usepackage{blindtext}
\usepackage{acronym}

\acrodef{AF}{Atrial Fibrillation}
\acrodef{SR}{Sinus Rhythm}
\acrodef{PVI}{Pulmonary Vein Isolation}
\acrodef{FIRM}{Focal Impulse and Rotor Modulation}
\acrodef{ECGI}{Electrocardiographic Imaging}
\acrodef{LAT}{Local Activation Time}
\acrodef{LA}{Left Atrium}
\acrodef{MRI}{Magnetic Resonance Imaging}
\acrodef{DT-MRI}{Diffusion Tensor Magnetic Resonance Imaging}
\acrodef{DE-MRI}{Delayed-Enhanced Magnetic Resonance Imaging}
\acrodef{LGE-MRI}{Late Gadolinium Enhancement Magnetic Resonance Imaging}
\acrodef{VT}{Ventricular Tachycardia}
\acrodef{CT}{Computed Tomography}
\acrodef{CV}{Conduction Velocity}
\acrodef{ICP}{Iterative Closest Point}
\acrodef{LSPV}{Left Superior Pulmonary Vein}
\acrodef{RSPV}{Right Superior Pulmonary Vein}
\acrodef{LIPV}{Left Inferior Pulmonary Vein}
\acrodef{RIPV}{Right Inferior Pulmonary Vein}
\acrodef{MV}{Mitral Valve}
\acrodef{BB}{Bachmann's Bundle}

\graphicspath{{figure/}}

\begin{document}

\title{Patient-Specific Heart Model Towards Atrial Fibrillation}

\author{Jiyue He, Arkady Pertsov, Sanjay Dixit, Katie Walsh, Eric Toolan, Rahul Mangharam
\thanks{Jiyue He, Rahul Mangharam: Department of Electrical and Systems Engineering, University of Pennsylvania, Philadelphia, PA, USA. jiyuehe@seas.upenn.edu. Arkady Pertsov: Department of Pharmacology, Upstate Medical University, Syracuse, USA. Sanjay Dixit, Katie Walsh: Department of Cardiac Electrophysiology, Hospital of the University of Pennsylvania, Philadelphia, PA, USA. Eric Toolan: Biosense Webster, Lansdowne, PA, USA. May 19, 2021.}}

\maketitle

\begin{abstract}
Atrial fibrillation is a heart rhythm disorder that affects tens of millions people worldwide. The most effective treatment is catheter ablation. This involves irreversible heating of abnormal cardiac tissue facilitated by electroanatomical mapping. However, it is difficult to consistently identify the triggers and sources that may initiate or perpetuate atrial fibrillation due to its chaotic behavior. We developed a patient-specific computational heart model that can accurately reproduce the activation patterns to help in localizing these triggers and sources. Our model has high spatial resolution, with whole-atrium temporal synchronous activity, and has patient-specific accurate electrophysiological activation patterns. A total of 15 patients data were processed: 8 in sinus rhythm, 6 in atrial flutter and 1 in atrial tachycardia. For resolution, the average simulation geometry voxel is a cube of 2.47 mm length. For synchrony, the model takes in about 1,500 local electrogram recordings, optimally fits parameters to the individual's atrium geometry and then generates whole-atrium activation patterns. For accuracy, the average local activation time error is 5.47 ms for sinus rhythm, 10.97 ms for flutter and tachycardia; and the average correlation is 0.95 for sinus rhythm, 0.81 for flutter and tachycardia. This promising result demonstrates our model is an effective building block in capturing more complex rhythms such as atrial fibrillation to guide physicians for effective ablation therapy.
\end{abstract}

\begin{IEEEkeywords}
medical cyber-physical systems, patient-specific, computational heart model, cardiac electrophysiology modeling, electroanatomical mapping, Mitchell-Schaeffer model
\end{IEEEkeywords}

\section{Introduction}
\begin{figure*}[t]
\centerline{\includegraphics[width = 1 \textwidth]{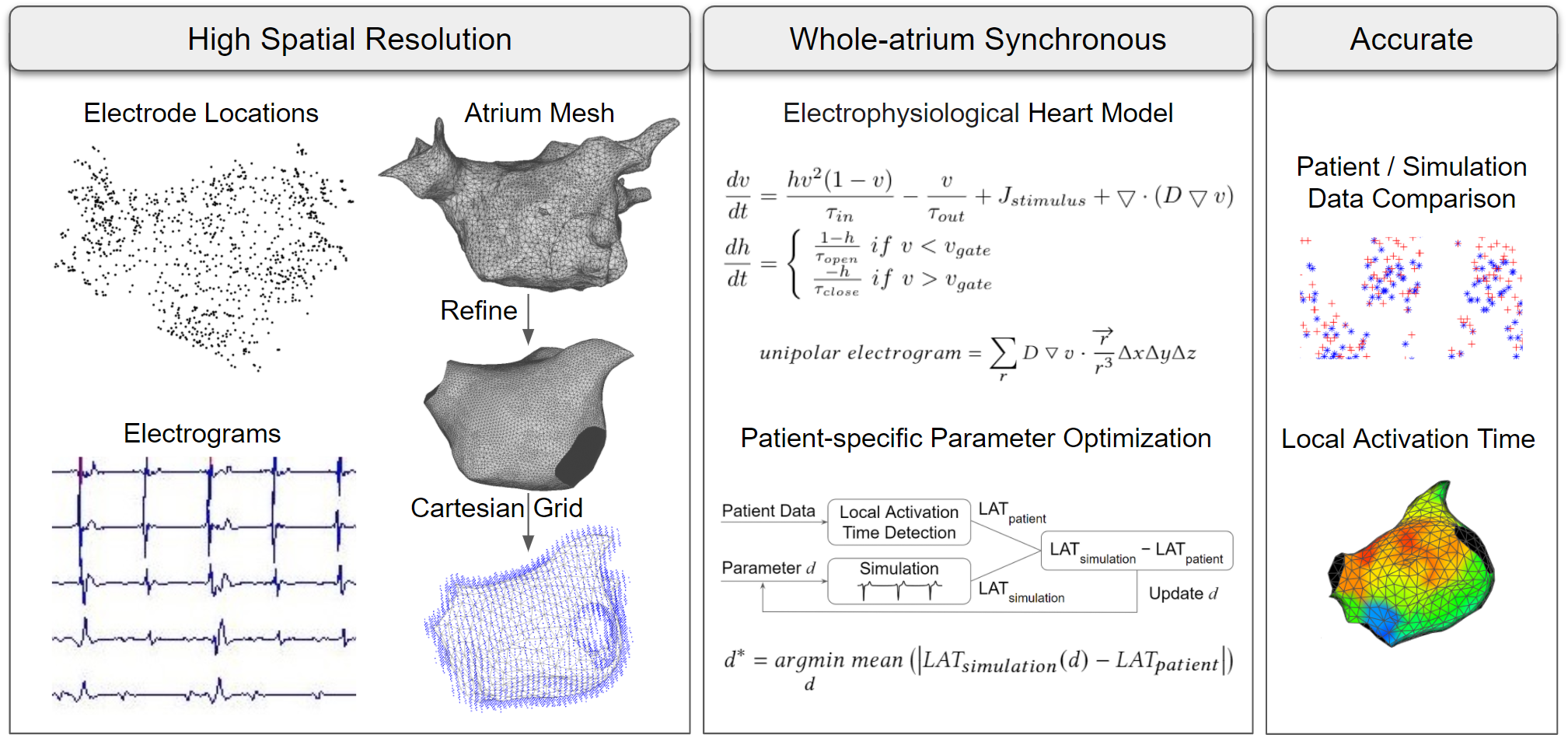}}
\caption{The overall process. Patient data consists of electrode locations, electrograms, and atrium 3D triangular mesh. Mesh is refined and pulmonary veins and mitral valve are cut out. Then, a Cartesian grid is created wrapping around the mesh. These data are input into the electrophysiological heart model, where parameters are fitted via an optimization process. Lastly, the heart model is validated through patient data.}
\label{fig:framework_overview}
\end{figure*}

\ac{AF} is a heart rhythm disorder where the normal beating in the two atria is irregular, and blood does not flow well to the ventricles. While the underlying mechanism of \ac{AF} is not clearly understood, it largely is due to the spatial variation in the conduction properties of the atrial myocardium. This causes the single wavefront propagating across the atria to being split into multiple wavefronts resulting in chaotic depolarization of the heart tissue and irregularly fast rhythm. \ac{AF} increases the risk of stroke and heart failure. If left untreated, it will become worse \cite{Wijffels}. 

One of the most effective treatments for \ac{AF} is catheter ablation. This involves irreversible radiofrequency heating of the sources that initiate or perpetuate \ac{AF}. The common current ablation protocol involves standard lesion locations for all patients. For example, in the \Ac{PVI} approach, ablation lesions encircle the pulmonary veins to prevent abnormal activations originated in the veins travel into the atria. 

Another common protocol involves \Ac{PVI} with additional ablation of non-pulmonary vein triggers and putative arrhythmia substrate \cite{Frankel}. While paroxysmal-\ac{AF} (i.e. spontaneous onset and termination) can be eliminated in 70-75\% of patients with a single ablation procedure, persistent-\ac{AF} can be eliminated in only about 50\% of patients with a single procedure \cite{AblationStats}. The reason is there are triggers other than pulmonary veins that cause fibrillation for the persistent-\ac{AF}. Electroanatomical mapping captures the heart geometry and tissue conductivity which are essential in identifying those trigger \cite{Santangeli}. 

The central challenge is to capture high-resolution synchronized electrograms across the entire atrium and analyze these data to identify potential triggers as ablation candidates to terminate \ac{AF}. We developed a high spatial resolution, temporally synchronous and patient-specific atrium model which captures electrophysiologic and anatomic parameters unique to each patient. 

Figure \ref{fig:framework_overview} shows the overall process of constructing the heart model. First, a high resolution electroanatomical map is exported from the Carto3 System, which contains atrium 3D triangular mesh, electrograms, and electrode locations. The mesh is processed to remove geometry defects, such as deep concave holes, intersecting triangular faces, and non-referenced vertices. The mitral valve and 4 pulmonary veins are cut out. Then a 3D Cartesian grid is generated and wrapped around the mesh for computing simulation. Then, these are put into the electrophysiological heart model, and patient-specific parameters are fitted via an optimization process. As a result of fitting parameters at every locations on the mesh, the heart model becomes whole-atrium synchronous. Lastly, the heart model is validated over 15 patients data by comparing \ac{LAT} between patient data and simulation data.

This model demonstrates the ability to accurately produce the activation patterns for an individual patient that can better identify \ac{AF} sources. 

\section{Related Work}

Several research groups had developed patient specific high resolution computational heart models \cite{Niederer}. Cabrera-Lozoya et al developed a model constructed from 3D \ac{DE-MRI} to simulate patient-specific post-infarction \ac{VT} abnormal electrograms. From \ac{DE-MRI} data, they segmented the ventricle into healthy myocardium, scar, and border zone. Then, three different sets of parameters were assigned to these three regions. They showed that their model can generate distinguishable electrograms for healthy region and scar region \cite{Cabrera}. Boyle et al developed a model for personalized ablation guidance of \ac{AF}. They obtained atrium geometry and fibrosis distribution from \ac{LGE-MRI}. Healthy regions and fibrosis regions are assigned with different parameters. Then virtual pacing are performed in simulation to identify all possible ablation targets \cite{Boyle}. Lim et al developed a model to do simulation-guided catheter ablation of \ac{AF}. Their model combined data from \ac{CT} images and electroanatomical maps to capture anatomy, fiber orientation, fibrosis, and electrophysiology. Different set of conduction parameters are assigned to fibrotic and non-fibrotic tissue \cite{Lim}. Corrado et al developed a model that had locally fitted parameters. They applied an S1-S2 electrical stimulation protocol from the coronary sinus and the high right atrium and recorded endocardium electrograms in the left atrium. The parameter fitting was done via a grid-search method that evaluated all the combinations of parameters within a range \cite{Corrado} \cite{Corrado2}. 

The major differences of our heart model are:
\begin{itemize}
\item Our model identifies tissue conduction and diffusion values locally at the voxel level, while other heart models assign two or three sets of fixed parameters to generic healthy or scar regions.
\item Our model fits the local parameters using self-activated electrograms rather then manual pacing-induced ones.
\item Our model does not use MRI or CT because they do not provide electrical activation data. Instead, we use electroanatomical mapping data which provides endocardium electrogram.
\item Our model is computationally light and will eventually be able to run during the clinical procedure for real-time ablation guidance.
\end{itemize}

It takes about 1 to 2 hours for Carto3 System to export 1 patient data. Our heart model program is implemented using Matlab (MathWorks), and runs on a laptop with 1 Intel Core i7 CPU. It takes about 2 minutes to read in a patient data to the computer. On average, it takes about 50 seconds to personalize one heart model. If our heart model was integrated into the Carto3 System, the step of export/import data would not be needed, which means the entire computation could be finished in 50 seconds. Further more, it could speed up hundreds times if GPU computing was implemented.

\section{High Spatial Resolution}
\subsection{Data Collection}

\begin{figure*}[htb]
\centerline{\includegraphics[width = 0.98\textwidth]{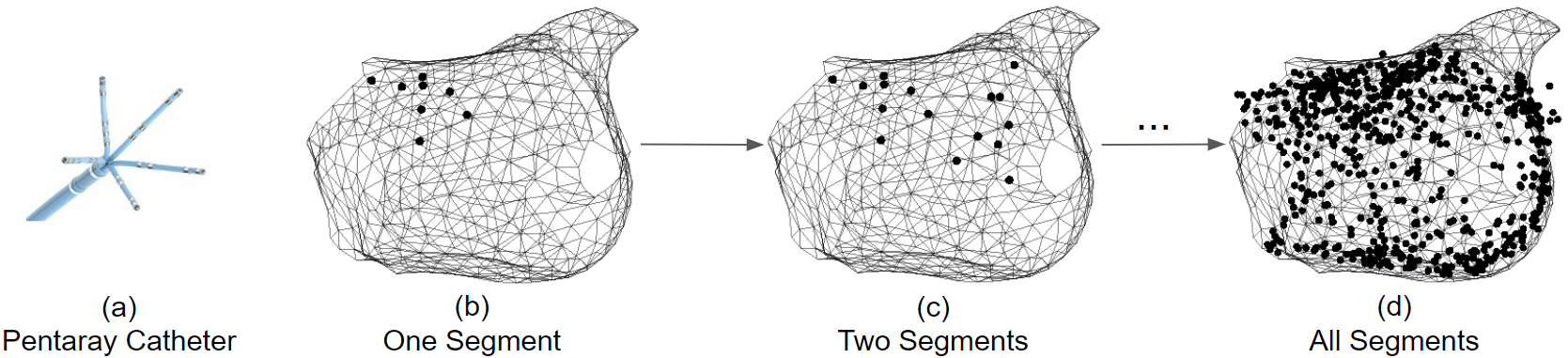}}
\caption{(a) Pentaray catheter is a star-shape catheter that has 20 electrodes. (b) Physicians will hold the catheter in a location for 2.5 seconds to obtain one segment of recording. (c) Then move the catheter to another location. (d) After 10 minutes, about 1,500 electrograms at different locations will be recorded.}
\label{fig:recording_segment}
\end{figure*}

\begin{figure}[htb]
\centerline{\includegraphics[width = 0.45\textwidth]{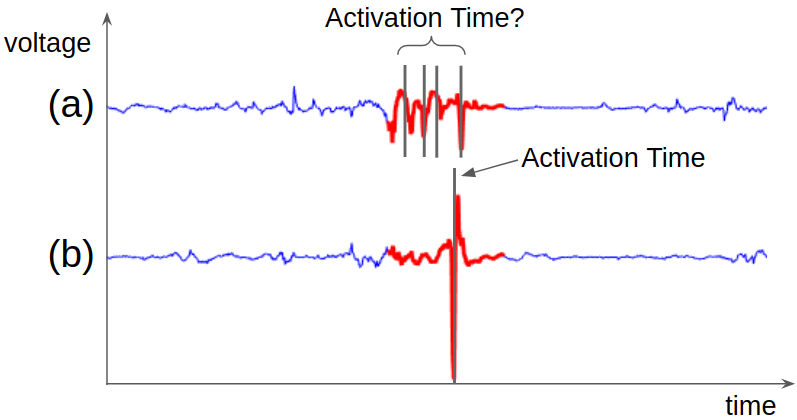}}
\caption{(a) Electrogram is too complex and fractionated, making it difficult to find the activation time. (b) Good electrogram. The activation time is easy to identify.}
\label{fig:fractionated_egm}
\end{figure}

At the start of the catheter ablation procedure, the physician captures an electroanatomical map with a roving mapping catheter from the Carto3 System at the Hospital of the University of Pennsylvania. As show in Figure \ref{fig:recording_segment}, it associates the 3D atrium triangular mesh with electrical activity at every location. The mesh mapping fill threshold was 5 mm, and electrogram recording filters were set at 2 to 240 Hz for unipolar electrograms, 16-500 Hz for bipolar electrograms, and 0.5-200 Hz for surface electrogram. Each map has about 1,500 electrogram recordings spread across the endocardium. Each electrogram records 2.5 seconds unipolar and bipolar signals at 1 kHz. Bad electrograms will introduce noise to the model, thus needed to be excluded: 

\begin{itemize}
\item Electrode distance to the nearest mesh vertex is greater than 8 mm. This is an empirical threshold we decided to use. The electrode was most likely not in contact with atrium tissue beyond this threshold.
\item Maximum voltage is less than 0.45 mV. These electrodes are either too far way from tissue or in contact with scar tissue \cite{He}. Neither will provide clean electrogram.
\item Complex and fractionated electrogram as shown in Figure \ref{fig:fractionated_egm}. Fractionated electrograms consist of multiple high frequency components with low amplitudes and long duration, makes it difficult to find the accurate activation time.
\end{itemize}

The amount of raw electrogram recordings captured and filtered are shown in Table \ref{tb:number_of_electrode_sample}. The percentage of good electrogram recordings is low for flutter and tachycardia maps because they contain more fractionated electrograms.

\begin{table}[htb]
\begin{center}
\caption{Number of Electrode Recordings}
\begin{tabular}{ c c c c c } 
\hline
ID & Rhythm & \# Electrode & \# Used & \% Used \\ \hline
1 & Sinus Rhythm & 976 & 557 & 57.1 \\
2 & Sinus Rhythm & 3263 & 1361 & 41.7 \\
3 & Sinus Rhythm & 3156 & 1788 & 56.7 \\
4 & Sinus Rhythm & 1488 & 663 & 44.6 \\
5 & Sinus Rhythm & 2477 & 1655 & 66.8 \\
6 & Sinus Rhythm & 2905 & 2079 & 71.6 \\
7 & Sinus Rhythm & 1744 & 861 & 49.4 \\
8 & Sinus Rhythm & 1801 & 1106 & 61.4 \\
\hline
\multicolumn{2}{ c }{Average} & 2226 & 1259 & 56.1 \\ \hline
\hline
9 & Flutter & 278 & 73 & 26.3 \\
10 & Flutter & 958 & 326 & 34.0 \\
11 & Tachycardia & 822 & 197 & 24.0 \\
12 & Flutter & 484 & 215 & 44.4 \\
13 & Flutter & 1227 & 198 & 16.1 \\
14 & Flutter & 714 & 130 & 18.2 \\
15 & Flutter & 1469 & 425 & 28.9 \\
\hline
\multicolumn{2}{ c }{Average} & 850 & 223 & 27.4 \\ \hline
\end{tabular}
\label{tb:number_of_electrode_sample}
\end{center}
\end{table}

\subsection{Triangular Mesh to Cartesian Grid}
\begin{figure}[htb]
\centering
\includegraphics[width = 0.4\textwidth]{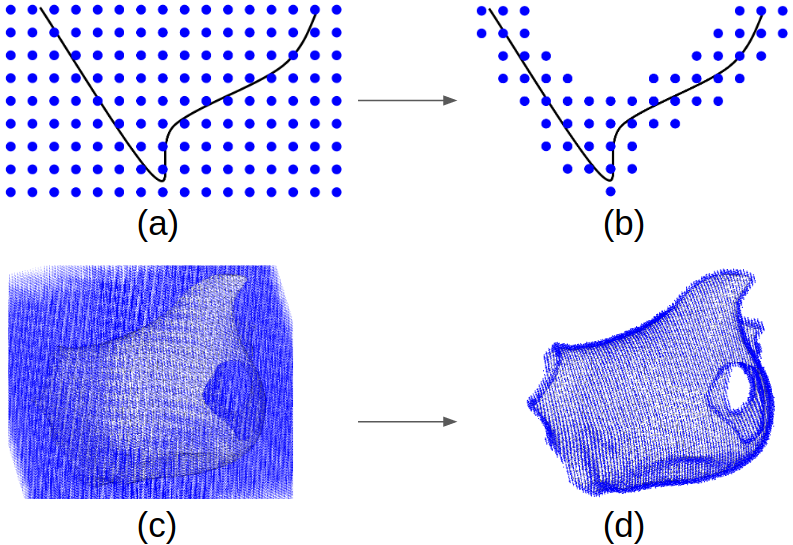}
\caption{Triangular mesh to Cartesian grid. (a) (b) Black line is atrium cross section, blue dots are Cartesian grid voxels. (a) A Cartesian grid encloses the entire atrium is created. (b) Voxels beyond a distance threshold to the mesh are removed. (c) (d) Show the same process on a patient's atrium.}
\label{fig:cartesian_grid}
\end{figure}

To achieve accurate activation wave propagation, we cut out the mitral valve and pulmonary veins. The cutting decision is based on atrium anatomy \cite{Gabriel}. The mesh is then refined to make each of the triangles the same size and shape. Mesh editing is done using a software called MeshLab \cite{meshlab}, and the two most helpful functionalities are 1) Simplification: quadric edge collapse decimation, can be used to smooth mesh imperfections; and 2) Remeshing: isotropic explicit remeshing, can be used to make triangles uniform in size and shape. For our heart model to simulate atrium action potential propagation, it requires a calculation of the second derivative of space. This is difficult to perform on triangular mesh because the vertices are not regularly positioned. In contrast, a Cartesian grid is regular. In Figure \ref{fig:cartesian_grid}, the black line represents a cross section of the atrium and the blue dots are Cartesian grid voxels. First, a Cartesian grid that encloses the entire atrium is created as shown in (a). Then the voxels that are beyond a distance threshold to the atrium mesh are removed as shown in (b). The distance threshold is equal to 1.5 times the average inter vertex distance. (c) (d) show the same process implemented on a patient's atrium mesh.

\begin{figure}[htb]
\centering
\includegraphics[width = 0.4\textwidth]{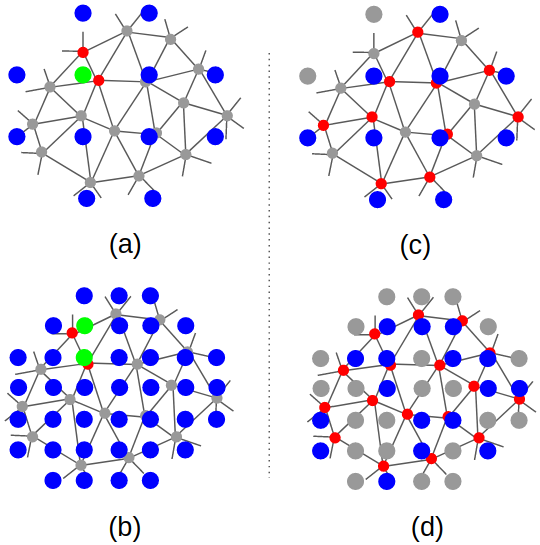}
\caption{The triangular mesh is atrium and the larger dots are Cartesian grid voxels. (a) The two red vertices are projected to the same green voxel. (b) The two red vertices are projected to two green voxels. (c) Red vertices received values, grey vertices did not. (d) All vertices received values from voxels.}
\label{fig:cartesian_grid_spacing}
\end{figure}

Patient electrograms are processed and the results are assigned to the nearest atrium mesh vertices. From atrium mesh vertices, values will be projected onto the nearest Cartesian grid voxels for simulation. Then the simulated results will be projected back to the nearest vertices of the atrium mesh. Denote $d_{voxel}$ the distance in between two neighboring Cartesian grid voxels, $d_{vertex}$ the average distance in between two neighboring atrium mesh vertices. It is important that this equation to be satisfied: $d_{voxel} < 1/2 \times d_{vertex}$. Figure \ref{fig:cartesian_grid_spacing} explains why. The triangular mesh is the atrium, and the larger dots are the Cartesian grid voxels. (a) If voxel spacing is too large, when projecting values from vertices to voxels, multiple vertices may be projected to the same voxel, causing values to be overwritten and information lost. Here the two red vertices are projected to the same green voxel. (b) If voxel spacing is small enough, each red vertex will be projected to a unique green voxel. (c) If voxel spacing is too large, when projecting values from voxels to vertices, some vertices may not receive any value. Here the red vertices received values, but the gray vertices do not. (d) If voxel spacing is small enough, all vertices will receive values. Notice that there may be voxels in (d) not being used as shown in gray, this does not create any problems, because the patient specific heart model is represented on the vertices. 

\ac{MRI} voxel resolution is 1 to 2 mm. While the mesh resolution can be easily edited by software MeshLab, we found that $d_{voxel} \approx 2$ mm will give satisfactory simulation accuracy. Detailed geometry resolution is shown in Table \ref{tb:resolution}.

\begin{table}[htb]
\begin{center}
\caption{Resolution of Our Heart Model}
\begin{tabular}{ c c c c c } 
\hline
ID & \# Vertices & $d_{vertex}$ (mm) & \# Voxels & $d_{voxel}$ (mm) \\ 
\hline
1 & 693 & 4.85 & 11807 & 2.43 \\
2 & 670 & 4.81 & 11583 & 2.41 \\
3 & 732 & 4.54 & 12486 & 2.27 \\
4 & 502 & 4.92 & 8640 & 2.46 \\
5 & 519 & 4.76 & 8975 & 2.38 \\
6 & 578 & 4.98 & 10031 & 2.49 \\
7 & 581 & 4.98 & 10207 & 2.49 \\
8 & 657 & 5.08 & 11339 & 2.54 \\
9 & 795 & 4.98 & 13973 & 2.49 \\
10 & 587 & 4.80 & 10251 & 2.40 \\
11 & 471 & 5.12 & 8097 & 2.56 \\
12 & 645 & 5.02 & 11116 & 2.51 \\
13 & 642 & 5.08 & 11134 & 2.54 \\
14 & 628 & 5.07 & 10960 & 2.54 \\
15 & 608 & 5.06 & 10637 & 2.53 \\
\hline
Average & 621 & 4.94 & 10749 & 2.47 \\ \hline
\end{tabular}
\label{tb:resolution}
\end{center}
\end{table}

\section{Whole-atrium Synchronous}
\subsection{Heart Model}
Our computational heart model implements the Mitchell-Schaeffer model \cite{Mitchell} as shown in Equation \ref{eq:mitchell_schaeffer}. It models the inward current ($hv^2(1-v) / \tau _{in}$) caused by sodium and calcium, and outward current ($-v / \tau_{out}$) caused by potassium, and external stimulus current ($J_{stimulus}$) of a cell. We use this model as its simplicity makes it efficient in 3D numerical simulations and model parameters provide direct insight into changes in electrical behavior. 

\begin{align}
\label{eq:mitchell_schaeffer}
\begin{split}
\frac{dv}{dt}&=\frac{hv^2(1-v)}{\tau _{in}}-\frac{v}{\tau _{out}}+J_{stimulus}+\bigtriangledown \cdot (D\bigtriangledown v)
\\
\frac{dh}{dt}&=\left\{\begin{matrix}\ \frac{1-h}{\tau_{open}} \ if\ v<v_{gate} \\ \ \frac{-h}{\tau_{close}} \ if\ v>v_{gate}\end{matrix}\right.
\end{split}
\end{align}

\begin{figure}[htbp]
\centering
\includegraphics[width = 0.25\textwidth]{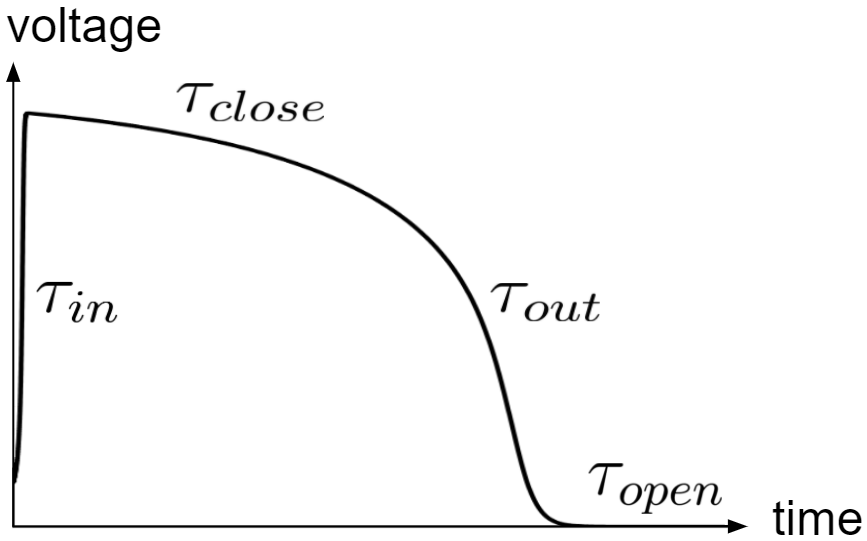}
\caption{The 4 $\tau$s influence the shape of action potential.}
\label{fig:tau}
\end{figure}

Detail of variables are as follows:

\begin{itemize}
\item $v$ is action potential voltage, scaled so that it ranges between 0 and 1. It may be scaled back to the original physiological values using the change of variables: $v' = v_{min} + v(v_{max} - v_{min})$. For example, $v_{min} = -70$ mV and $v_{max} = 30$ mV.
\item $h$ is the gating variable, varies between 0 and 1.
\item $J_{stimulus}$ is an external current applied in brief pulses, it initiates local activation.
\item $\tau_{in}$, $\tau_{close}$, $\tau_{out}$, and $\tau_{open}$ are the parameters controlling the action potential shape as shown in Figure \ref{fig:tau}. Note that by changing any one of them will change the entire shape of action potential, just that the main effect of change is as shown.
\item $v_{gate}$ is change-over voltage.
\item $\bigtriangledown \cdot (D\bigtriangledown v)$ is diffusion term that contributes action potential propagation, where $\bigtriangledown \cdot$ is the divergence operator and $\bigtriangledown$ is the gradient operator.
\item $D$ is the diffusion tensor that controls the activation wave speed and direction. This is the most important and relevant parameter for this paper.
\end{itemize}

\begin{figure}[htbp]
\centering
\includegraphics[width = 0.25\textwidth]{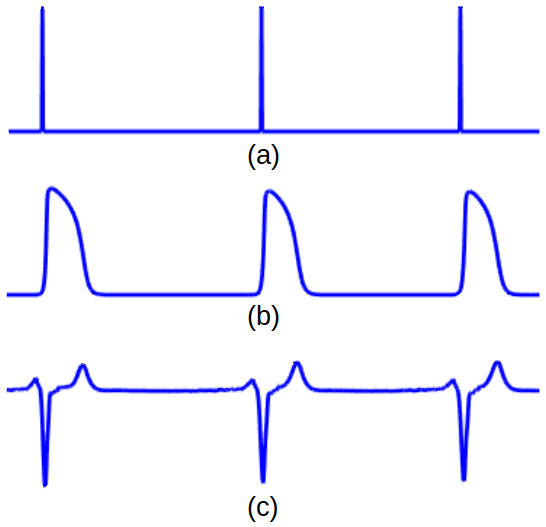}
\caption{(a) The stimulus impulse train $J_{stimulus}$. (b) Action potential. (c) Unipolar electrogram.}
\label{fig:J_stimulus}
\end{figure}

The stimulus current term $J_{stimulus}$ represents the initiation of local activation. For a normal heart, the sinus node in the right atrium and the Bachmann bundle region in the left atrium will have $J_{stimulus} \neq 0$, and for other locations, $J_{stimulus}=0$; for an abnormal heart having \ac{AF}, $J_{stimulus}$ can be used to simulate focal sources. If a vertex on the mesh is a pacing site, then $J_{stimulus}$ is equal to an impulse train as shown in Figure \ref{fig:J_stimulus} (a).

To compute unipolar electrogram from action potential $v$, the dipole Equation \ref{eq:electrogram} is implemented, note that the dot in the middle of the equation is a dot product. Unipolar electrogram is obtained by integrating over the entire Cartesian grid. 

\begin{equation}
\label{eq:electrogram}
unipolar\ electrogram=\sum_{r}^{ } D\bigtriangledown v \cdot \frac{\overrightarrow{r}}{r^{3}} \Delta x\Delta y\Delta z
\end{equation}

Detail of variables are as follows:

\begin{itemize}
\item $r$ is the distance from the point of interest on the triangular mesh to the other point in the Cartesian grid. 
\item $\Delta x$, $\Delta y$, and $\Delta z$ are distances of neighboring voxels in the x, y, and z direction. 
\end{itemize}

In practice, if $r<1$ mm, we assign $r=1$ mm to avoid the term $1/r^3$ becoming too large. The physical meaning of doing so is: the electrode is at least 1 mm away from tissue.

\subsection{Equations in Discrete Form}

To program Equation \ref{eq:mitchell_schaeffer}, we transform it into discrete form. To simplify computation, we do not consider fiber orientations, and assume isotropic diffusion. Therefore, the diffusion tensor $D$ simplifies to a diagonal matrix with 3 diagonal elements all equal to $d(x,y,z)$. Also note that $v = v(x,y,z)$. Therefore the diffusion term is transformed into Equation \ref{eq:diffusion_term_discrete_form}. In Equation \ref{eq:derivatives}, $v(x+\Delta x,y,z)$ is the neighbor of $v(x,y,z)$ in the positive x direction.

\begin{align}
\label{eq:diffusion_term_discrete_form}
\begin{split}
&\bigtriangledown \cdot \left ( D\bigtriangledown v \right ) = \bigtriangledown \cdot \left ( \begin{bmatrix} d & 0 & 0\\ 0 & d & 0\\ 0 & 0 & d\end{bmatrix} \begin{bmatrix} \frac{\partial v}{\partial x}\\ \frac{\partial v}{\partial y}\\ \frac{\partial v}{\partial z}
\end{bmatrix}\right )
\\
&=\frac{\partial (d\frac{\partial v}{\partial x})}{\partial x} + \frac{\partial (d\frac{\partial v}{\partial y})}{\partial y} + \frac{\partial (d\frac{\partial v}{\partial z})}{\partial z}
\\
&=\frac{\partial d}{\partial x}\frac{\partial v}{\partial x}+d\frac{\partial^{2} v}{\partial x^{2}}  +  \frac{\partial d}{\partial y}\frac{\partial v}{\partial y}+d\frac{\partial^{2} v}{\partial y^{2}} + \frac{\partial d}{\partial z}\frac{\partial v}{\partial z}+d\frac{\partial^{2} v}{\partial z^{2}}
\end{split}
\end{align}

Where

\begin{align}
\label{eq:derivatives}
\begin{split}
\frac{\partial v}{\partial x} &= \frac{v(x+\Delta x,y,z) - v(x-\Delta x,y,z)}{2 \Delta x} \\
\frac{\partial^2 v}{\partial x^2}&= \frac{v(x+\Delta x,y,z)-2v(x,y,z)+v(x-\Delta x,y,z)}{(\Delta x)^2}
\end{split}
\end{align}

To compute the next time step, we implemented Equation \ref{eq:discrete_mitchell_schaeffer}. Here $\Delta t=0.1$ ms is the simulation time step. Similar equations for the variable $h$.

\begin{align}
\label{eq:discrete_mitchell_schaeffer}
\begin{split}
&\frac{dv}{dt} = (v(t+\Delta t)-v(t))/\Delta t = f(v(t)) \\
&\Rightarrow v(t+\Delta t) = f(v(t)) \Delta t + v(t)
\end{split}
\end{align}

\subsection{Boundary Condition}

To properly simulate atrium activation wave, the no-flux boundary condition is applied to the Cartesian grid voxels, written in discrete form in Equation \ref{eq:boundary_condition}. Similar equations for y and z axes. 

\begin{align}
\label{eq:boundary_condition}
\begin{split}
&\frac{\partial v(x,y,z)}{\partial x} = \frac{v(x+\Delta x,y,z) - v(x-\Delta x,y,z)}{2 \Delta x}=0 \\
&\Rightarrow v(x+\Delta x,y,z) = v(x-\Delta x,y,z)
\end{split}
\end{align}

\begin{figure}[htb]
\centering
\includegraphics[width = 0.40\textwidth]{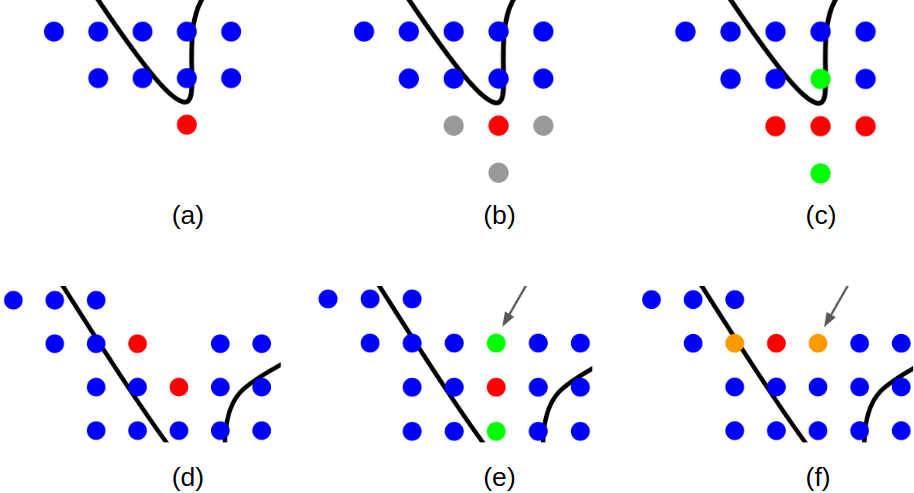}
\caption{Boundary conditions explained in 2D. The black line is a cross section of atrium mesh. The dots are Cartesian grid voxels. (a) (b) (c) Apply boundary condition to a convex corner voxel (red). (b) Create 3 dummy voxels. (c) The -x and +x dummy voxels copy the value of the red one, and the -y voxel copy the value of the +y voxel (green). (d) (e) (f) Apply boundary condition to concave corner voxels (red). (e) Create a dummy voxel (pointed by the arrow) and copy the value of the -y voxel (green). (f) Create a dummy voxel (pointed by the arrow) and copy the value of the -x voxel (orange).}
\label{fig:boundary_condition}
\end{figure}

A voxel is a boundary voxel if it does not have 6 neighbors (+x, -x, +y, -y, +z, and -z directions). Figure \ref{fig:boundary_condition} explains how to implement the boundary condition stated in Equation \ref{eq:boundary_condition}. The black line is a cross section of atrium mesh. The dots are Cartesian grid voxels. (a) (b) (c) explain a convex corner case: (b) create 3 dummy voxels (grey); (c) -x and +x dummy voxels copy the value of the red, -y dummy voxel copy the value of the +y voxel (green). (d) (e) (f) explain a concave corner case: (e) this red voxel does not have a neighbor in +y direction, thus a dummy voxel is created (pointed by the arrow) there and copies the value of the -y voxel (green); (f) this red voxel does not have a neighbor in the +x direction, thus a dummy voxel is created (pointed by the arrow) there and copies the value of the -x voxel (orange). Note that the dummy voxels in (e) and (f) are at the same location, but depends on which red voxel is the boundary, this dummy voxel copies different neighbor voxel value, and it will create small errors in solving the equations.

\subsection{Patient-specific Parameter Optimization}                                        
\begin{figure}[htb]
\centering
\includegraphics[width = 0.30\textwidth]{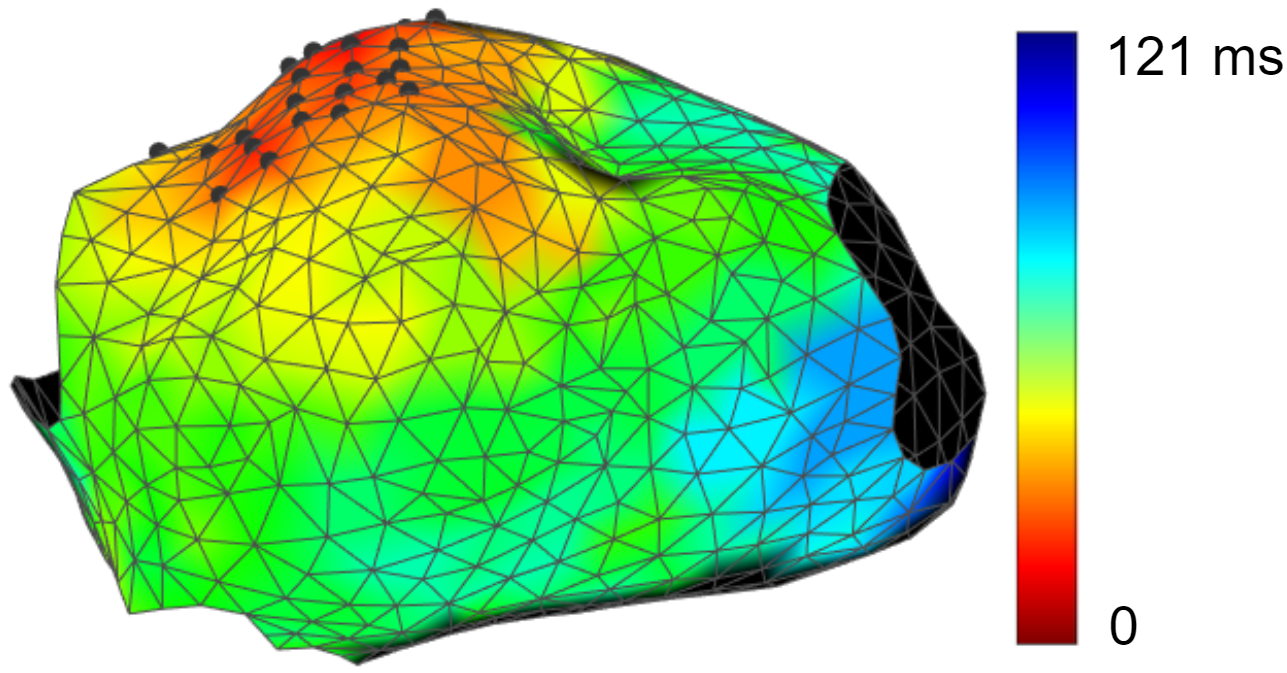}
\caption{\ac{LAT} map. Red represents early and blue represents late activation. Vertices marked in black are the pacing sites.}
\label{fig:active_sites}
\end{figure}

The goal is to match the simulated and patient \ac{LAT}. Each electroanatomical map contains about 1,500 electrogram recordings (each of 2.5 seconds) spread across the endocardium. As these recordings are taken at different times, they do not provide synchronous view of the whole atrium. \ac{SR}, flutter, and tachycardia are periodic rhythms, thus the time asynchronous problem can be solved by aligning them to a reference channel. The reference channel can be one of the surface electrodes or one of the coronary sinus catheter electrodes. Once the reference channel is picked, it will remain the same for all electrograms within the map. Color coding each electrode's activation time will result in a \ac{LAT} map as shown in Figure \ref{fig:active_sites}, where red represents early activation, blue represents late activation, so that activation waves travel from early sites to late sites. Patient's bipolar electrograms are chosen to detect \ac{LAT}, since bipolar recordings are less noisy than unipolar recordings. \ac{LAT} is found at the time instance of maximum absolute slope computed with Equation \ref{eq:LAT_bipolar}. Simulated \ac{LAT} is detected from simulated unipolar electrograms as the time instance of maximum negative slop computed with Equation \ref{eq:LAT_unipolar} \cite{Cantwell}. 

\begin{equation}
\label{eq:LAT_bipolar}
LAT_{patient} = \underset{t}{argmax}\left ( \left | \frac{dV_{bipolar}(t)}{dt} \right | \right )
\end{equation}

\begin{equation}
\label{eq:LAT_unipolar}
LAT_{simulation} = \underset{t}{argmax}\left ( - \frac{dV_{unipolar}(t)}{dt} \right )
\end{equation}

Every vertex of the atrium triangular mesh has 6 parameters: $\tau_{in}$, $\tau_{out}$, $\tau_{open}$, $\tau_{close}$, $v_{gate}$, and $d$. Since the first 5 parameters mainly affect action potential shape, and for \ac{SR}, flutter and tachycardia, the \ac{LAT} is mostly influenced by the timing of the up stroke of action potential. Therefore, we only need to tune the diffusion coefficient $d$, which controls the activation wave propagation speed. For the other 5 parameters, we assign them nominal values for all vertices of the atrium mesh as shown in Table \ref{tb:nominal_value} \cite{Cabrera}.

\begin{table}[htb]
\begin{center}
\caption{Action Potential Parameters Nominal Values}
\begin{tabular}{ c c c c c } 
\hline
$\tau_{in}$  & $\tau_{out}$ & $\tau_{open}$ & $\tau_{close}$ & $v_{gate}$ \\ 
\hline
0.3 & 6 & 120 & 150 & 0.13 \\ 
\hline
\end{tabular}
\label{tb:nominal_value}
\end{center}
\end{table}

\begin{figure}[ht]
\centering
\includegraphics[width = 0.45\textwidth]{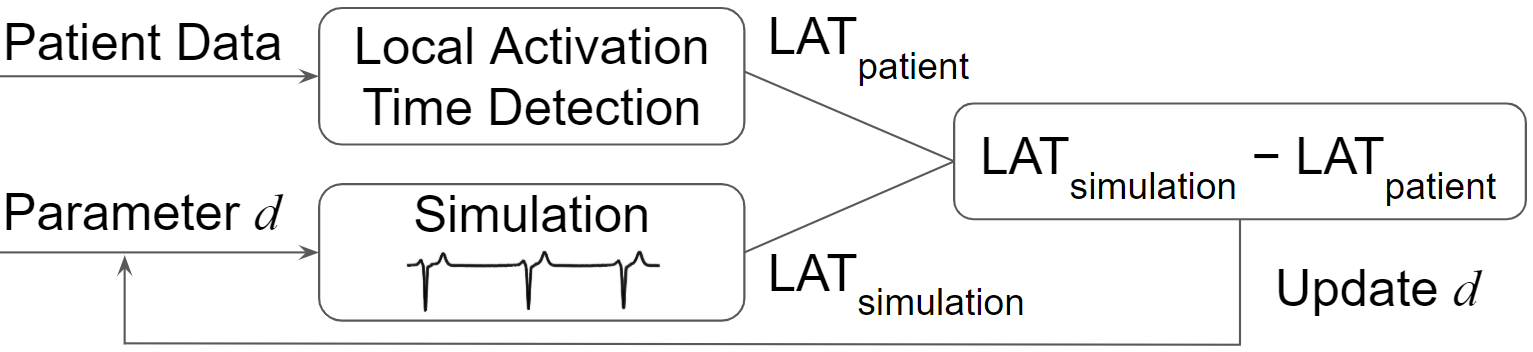}
\caption{Parameter optimization. $d$ values are updated according to the errors.}
\label{fig:estimation}
\end{figure}

The process of optimizing $d$ is shown in Equation \ref{eq:empirical_risk_minimization} and Figure \ref{fig:estimation}. $d$ is a $N\times1$ vector, where $N$ is the amount of data samples. $d$ is optimized when the \ac{LAT} error between patient data and simulation data is minimized. The optimization is an iterative process, during the iterations, $d$ is updated according to Equation \ref{eq:update_d}, here $\varepsilon = 0.01$. $LAT_{simulation,n}-LAT_{patient,n} > 0$ means for the $n$-th vertex, the simulation activation wave appeared later than patient activation wave, therefore need to increase its diffusion coefficient to speed up activation wave propagation, and the increase amount is $\left ( LAT_{simulation,n}-LAT_{patient,n} \right ) \varepsilon$. If the opposite happened, $LAT_{simulation,n}-LAT_{patient,n} < 0$ would decrease the diffusion coefficient. There are 2 constrains: 1) $d \geq 0.001$, 2) if the value is greater than 3 standard deviations of the neighboring values, replace it with the average of neighbor values. The optimization algorithm stopping criterion is as follows: 1) number of iterations greater than 200, or 2) maximum value of action potential is abnormal (greater than 10), or 3) the 6 previous \ac{LAT} errors are greater than the minimum \ac{LAT} error.

\begin{equation}
\label{eq:empirical_risk_minimization}
d^*=\underset{d}{argmin}\ mean\left( \left| LAT_{simulation}(d)-LAT_{patient} \right| \right)
\end{equation}

\begin{equation}
\label{eq:update_d}
d_{new} = d_{old} + \left ( LAT_{simulation}-LAT_{patient} \right ) \varepsilon
\end{equation}

Because the heart model is a highly nonlinear system, initial value of $d$ can affect its optimum value, therefore, the resulting optimum values may not be globally optimum. We had tried using different initial values, however, some lead to unstable simulation, and some lead to large errors. From trials and errors, an empirical good guess for the initial value is shown in Equation \ref{eq:empirical_diffusion_coefficient}. The advantage of this equation is that it gives a value that is adaptive to the heart model geometry's resolution. For example, if the voxel spacing of the Cartesian grid of the atrium is $\Delta x=\Delta y=\Delta z=2.5$ mm, and the simulation time step is $\Delta t=0.1$ ms, then $0.05 \times 2.5 / 0.1 = 1.25$ is a good initial value.

\begin{equation}
\label{eq:empirical_diffusion_coefficient}
d_{initial} = 0.05 \times \Delta x / \Delta t
\end{equation}

Besides optimizing $d$, another important term in the heart model is the stimulus current $J_{stimulus}$. To specify $J_{stimulus}$, first, patient data are processed to obtain \ac{LAT} of every electrode locations; then the earliest activation location is found as it is the vertex that has the smallest \ac{LAT} value; lastly, vertices having \ac{LAT} no larger than 20 ms of the earliest one are also found. These vertices are marked in black in Figure \ref{fig:active_sites}, which are the pacing sites; we assigned $J_{stimulus}$ brief impulse of magnitude 10 and duration $2 \Delta t$ to the pacing sites at the time instance according to their \ac{LAT} values; all other vertices are not pacing sites, and are assigned $J_{stimulus}=0$.

\section{Model Accuracy}
\begin{table*}[ht]
\begin{center}
\caption{Performance Comparison}
\begin{tabular}{ c c | c c c c | c c c c | c c c c } 
\hline
\multicolumn{2}{c}{} & \multicolumn{4}{|c}{Model I: $d$ fitted individually} & \multicolumn{4}{|c}{Model II: $d$ fitted uniformly} & \multicolumn{4}{|c}{Model III: different $\tau$s}\\
\hline
ID & Rhythm & LAT Err & RMSE & Corr & Acc & LAT Err & RMSE & Corr & Acc & LAT Err & RMSE & Corr & Acc\\
\hline
1 & Sinus Rhythm & 7.66 & 9.76 & 0.94 & 93.16 & 8.73 & 11.06 & 0.91 & 92.21 & 8.45 & 10.73 & 0.92 & 92.45 \\ 
2 & Sinus Rhythm & 5.74 & 8.19 & 0.96 & 95.25 & 10.71 & 14.40 & 0.88 & 91.15 & 9.28 & 12.05 & 0.93 & 92.33 \\
3 & Sinus Rhythm & 5.86 & 8.64 & 0.92 & 92.85 & 7.15 & 10.08 & 0.85 & 91.28 & 5.41 & 7.61 & 0.92 & 93.4 \\ 
4 & Sinus Rhythm & 5.22 & 7.99 & 0.95 & 94.62 & 5.50 & 8.55 & 0.93 & 94.33 & 4.89 & 7.04 & 0.95 & 94.96 \\
5 & Sinus Rhythm & 4.42 & 6.13 & 0.94 & 93.77 & 5.74 & 7.79 & 0.89 & 91.92 & 5.17 & 6.90 & 0.93 & 92.72 \\
6 & Sinus Rhythm & 5.65 & 7.81 & 0.95 & 93.36 & 6.19 & 8.19 & 0.91 & 92.72 & 5.52 & 7.24 & 0.94 & 93.50 \\
7 & Sinus Rhythm & 4.48 & 6.61 & 0.96 & 94.60 & 4.61 & 6.76 & 0.95 & 94.45 & 4.04 & 5.94 & 0.96 & 95.13 \\
8 & Sinus Rhythm & 4.69 & 6.81 & 0.96 & 95.70 & 7.50 & 9.97 & 0.91 & 93.12 & 6.67 & 8.49 & 0.95 & 93.88 \\
\hline
\multicolumn{2}{ c| }{Average} & 5.47 & 7.74 & 0.95 & 94.16 & 7.02 & 9.60 & 0.90 & 92.65 & 6.18 & 8.25 & 0.94 & 93.55 \\ \hline
\hline
9 & Flutter & 8.34 & 16.40 & 0.91 & 93.05 & 12.91 & 17.49 & 0.89 & 89.24 & 5.98 & 9.66 & 0.97 & 95.01 \\
10 & Flutter & 6.87 & 11.82 & 0.83 & 93.87 & 12.22 & 16.85 & 0.67 & 89.09 & 6.88 & 10.79 & 0.87 & 93.86 \\
11 & Tachycardia & 5.49 & 10.11 & 0.85 & 93.76 & 9.95 & 15.24 & 0.62 & 88.69 & 5.65 & 11.23 & 0.84 & 93.58 \\
12 & Flutter & 13.77 & 27.96 & 0.87 & 92.31 & 22.37 & 31.12 & 0.82 & 87.50 & 10.07 & 15.59 & 0.96 & 94.38 \\
13 & Flutter & 10.89 & 23.42 & 0.50 & 92.39 & 13.09 & 21.68 & 0.58 & 90.85 & 8.66 & 17.35 & 0.74 & 93.94 \\
14 & Flutter & 17.71 & 23.45 & 0.79 & 91.01 & 17.81 & 24.20 & 0.69 & 90.96 & 13.11 & 17.23 & 0.84 & 93.34 \\
15 & Flutter & 13.70 & 21.49 & 0.91 & 92.75 & 22.22 & 32.18 & 0.84 & 88.24 & 10.53 & 17.56 & 0.94 & 94.43 \\
\hline
\multicolumn{2}{ c| }{Average} & 10.97 & 19.24 & 0.81 & 92.73 & 15.80 & 22.68 & 0.73 & 89.22 & 8.70 & 14.20 & 0.88 & 94.08 \\ \hline
\end{tabular}
\label{tb:performance_summary}
\end{center}
\begin{flushleft}
LAT Err: local activation time error, defined in Equation \ref{eq:lat_diff}, unit: ms. RMSE: root-mean-square error, unit: ms. Corr: correlation. Acc: accuracy, defined in Equation \ref{eq:accuracy}.
\end{flushleft}
\end{table*}

A total of 15 patients data were processed, including 8 sinus rhythm, 6 atrial flutter, and 1 atrial tachycardia map. Directly compare our heart model to other research groups' heart models is not easy, because different research groups validate their models using different methods; also different models are built upon different inputs: some require MRI data, some require optical mapping. It is not realistic to acquire that many kinds of data from our patients. However, we attempted a comparison, and results are shown in Model I, II and III in Table \ref{tb:performance_summary}. In the table, correlation is a statistic that measures linear correlation between two variables. It has a value in the range of -1 to 1. A value of 1 is total positive linear correlation, 0 is no linear correlation, and -1 is total negative linear correlation. The denominator in Equation \ref{eq:accuracy} is the range of patient \ac{LAT}, its physical meaning is: the time it takes the activation wave to travel the entire atrium. 

\begin{equation}
\label{eq:lat_diff}
\textit{LAT Err} = \frac{1}{N}\sum_{n=1}^{N}\left | LAT_{simulation,n}-LAT_{patient,n} \right |
\end{equation}

\begin{equation}
\label{eq:accuracy}
Accuracy = \left ( 1-\frac{\textit{LAT Err}}{range(LAT_{patient})}\right ) \times 100\%
\end{equation}

Model I is the performance of our heart model, where the diffusion coefficients $d$ are fitted for each individual vertex of the patient's atrium mesh. \ac{LAT} maps and patient v.s. simulation plots are provided in the appendix. Model II is the performance of which the diffusion coefficients $d$ are fitted uniformly: having the same value for all vertices. Compare to Model I, performance of Model II is worse: the average \ac{LAT} error increases, root-mean-square error increases, correlation decreases, and accuracy decreases. Model III is the performance of our heart model, but with a different set of action potential parameters. A comparison of Model I and III is described in Discussion section.

\section{Discussion}
\subsection{Modeling Atrial Fibrillation}

Our patient-specific high-resolution computational heart model is capable of reproducing patient specific whole chamber electrophysiology behaviors such as zigzag propagation and rotors. These phenomena depict \ac{AF} behaviors where a single wavefront propagating across the atria splits into multiple wavefronts resulting in chaotic depolarization of the heart tissue. 

\begin{figure}[htbp]
\centering
\includegraphics[width = 0.4\textwidth]{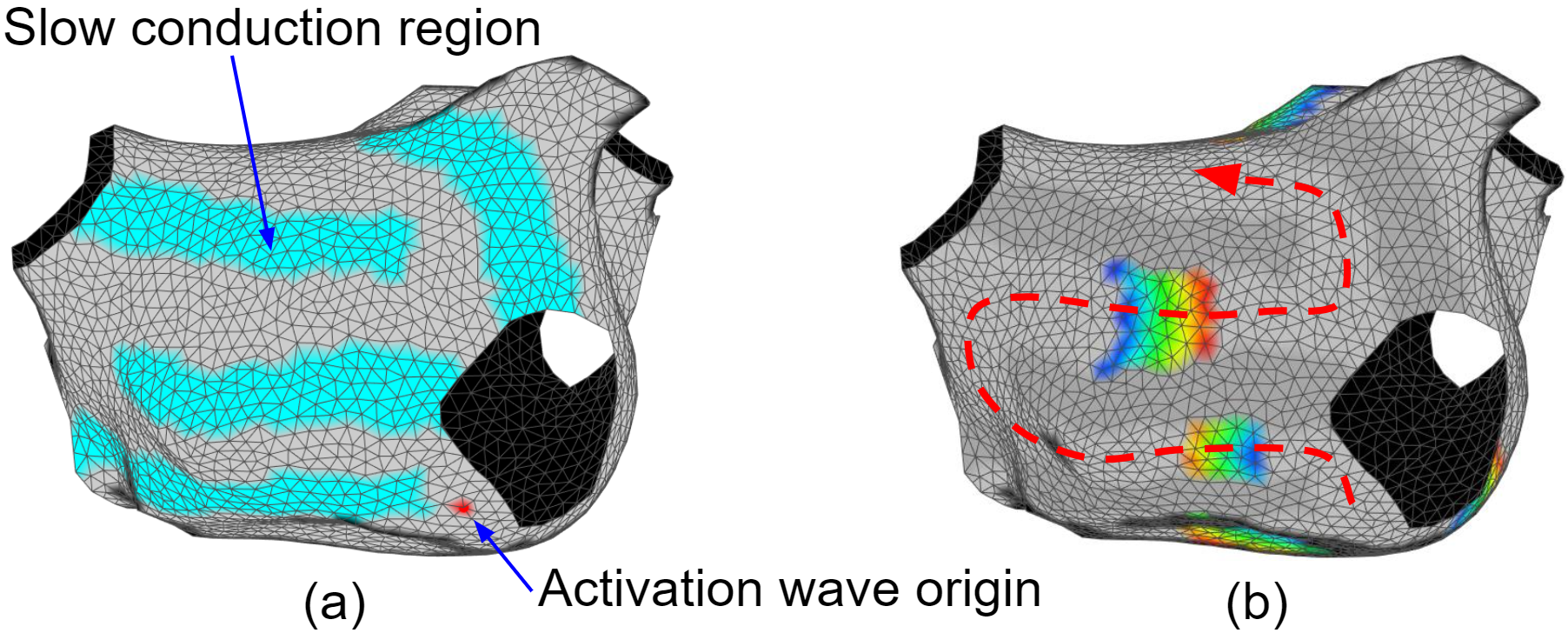}
\caption{Zigzag propagation. (a) Simulation setup: slow conduction regions are colored in cyan, activation origin is assigned to the red dot. (b) Slow conduction regions act as activation wave propagation barriers, forcing the wave travel in a zigzag manner as shown in the red dashed line.}
\label{fig:zigzag}
\end{figure}

Zigzag propagation is an explanation of slow conduction in the infarcted heart \cite{Bakker}. Figure \ref{fig:zigzag} (a) shows the simulation setup. Cyan regions are set as infarcted, where the conduction coefficients are set to 100 times smaller than healthy regions in gray. An activation wave origin is set at the lower right corner shown as a red dot. (b) shows a static time frame of the activation wave simulation. The wave front is colored in red, and the wave tail is in blue. The infarcted regions act as wave propagation blockage, forcing the wave travel in a zigzag manner as shown with the red dash line, which effectively increases the distance the wave needs to go through to reach the upper region. The result is that the effective wave propagation speed in the up-down direction is slower than the left-right direction. Being able to identify different conductivity regions is helpful for guiding catheter ablation. Slow conduction indicates scar regions, where arrhythmia is more likely to be formed, thus physician may only need to make ablations that connect scar regions to stop arrhythmia.

\begin{figure}[htbp]
\centering
\includegraphics[width = 0.4\textwidth]{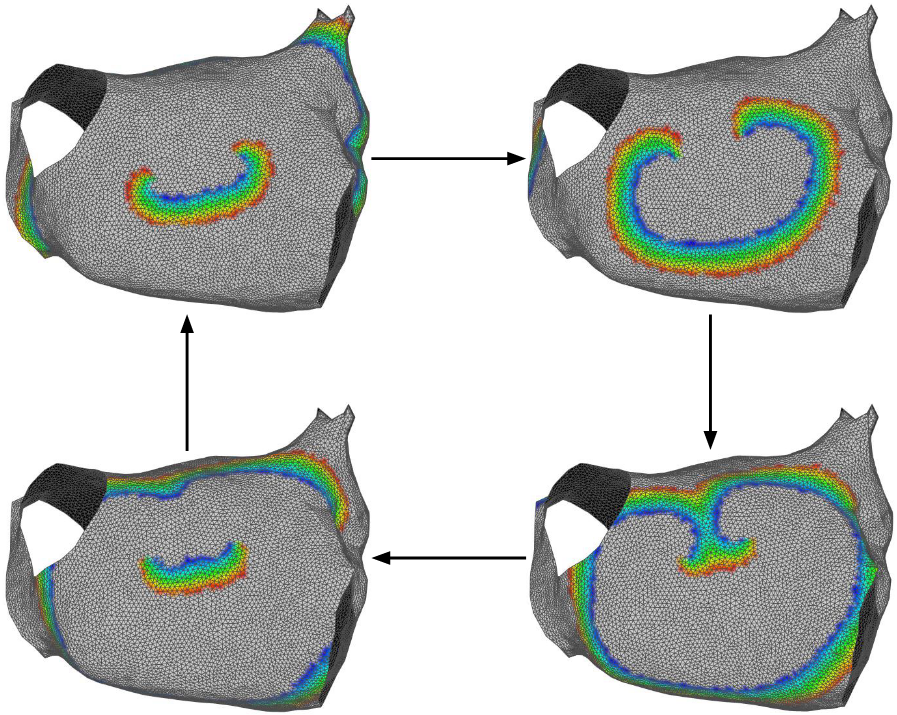}
\caption{A rotor is a rotating activation wave that goes around a meandering pivot point repeatedly, causing abnormal heart rhythm.}
\label{fig:rotor}
\end{figure}

Figure \ref{fig:rotor} shows simulated rotors. Rotors are rotating waves that spin fast, last long and will cause fast activations, which is one of the main AF characteristics. Rotor origins are common ablation targets.

\subsection{Importance of Action Potential Parameters}

For \ac{AF}, tissue heterogeneity is more pronounced. In this paper, we assumed tissue conductivity equal in x, y, and z direction. Relaxing this assumption will allow a more accurate heart model, such as incorporating fiber orientations. Also, having patient-specific action potential parameters are also important. To illustrate this, we ran our program with a different set of action potential parameters: $\tau_{in} = 0.3$, $\tau_{out} = 3$, $\tau_{open} = 120$, $\tau_{close} = 50$, $v_{gate} = 0.13$ (instead of $\tau_{in} = 0.3$, $\tau_{out} = 6$, $\tau_{open} = 120$, $\tau_{close} = 150$, $v_{gate} = 0.13$). The resulting performance is shown in Table \ref{tb:performance_summary} Model III. Compare to Model I, the average performance of \ac{SR} decreases 12.98\% ((5.47-3.97)/5.47), but the average performance of flutter and tachycardia increases 20.69\% ((10.97-8.7)/10.97). The reason may be that the original parameters are more accurate for \ac{SR} maps while these parameters are more accurate for flutter and tachycardia maps.

\subsection{Importance of having Multiple Pacing Sites}
The patient data we processed mostly have only one activation origin. To develop a robust heart model, fitting data that contains multiple activation sites is necessary. 

\subsection{Limitations on Flutter and Tachycardia}

\begin{figure}[htbp]
\centering
\includegraphics[width = 0.45\textwidth]{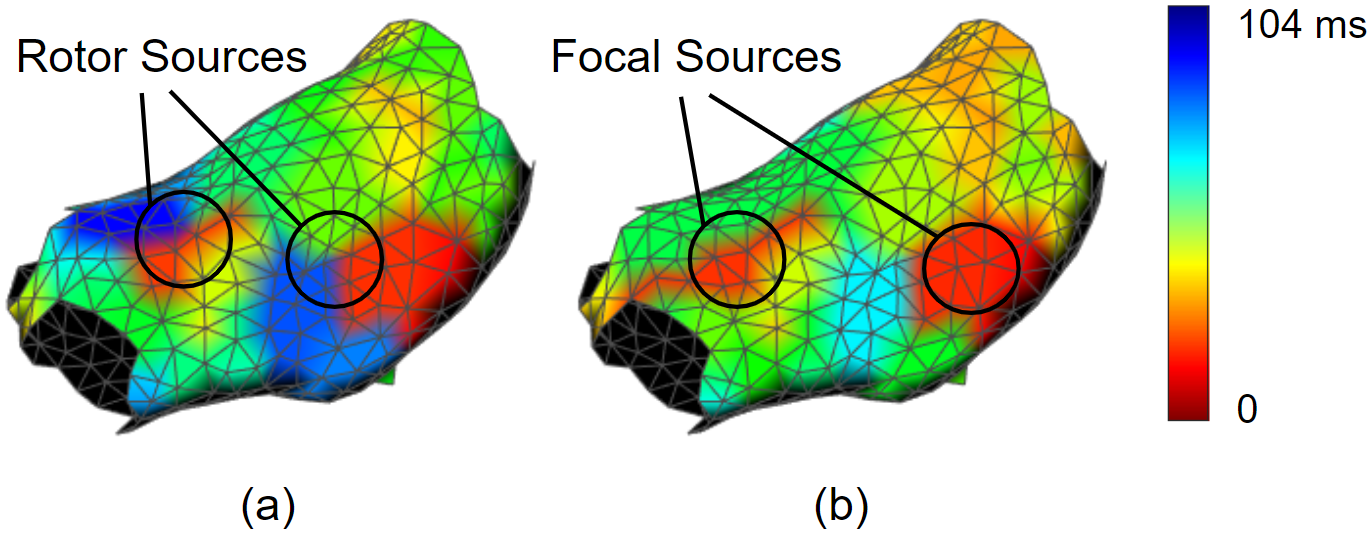}
\caption{The simulated electrophysiologic behavior did not match patient data. (a) Patient \ac{LAT} map. (b) Simulation \ac{LAT} map.}
\label{fig:tachycardia}
\end{figure}

In Table \ref{tb:performance_summary} Model I, the \ac{LAT} fitting is quite good for the atrial tachycardia patient (ID 11): having a \ac{LAT} error of 5.49 ms, patient and simulation \ac{LAT} correlation of 0.85. But upon a closer look at their \ac{LAT} map shown in Figure \ref{fig:tachycardia}, there is a problem: the two rotor sources were simulated by two focal sources. Obviously the activation wave dynamics of rotor sources and focal sources are different, thus our heart model did not correctly reproduced this patient's electrophysiologic behaviors.

Lombardo et al fitted their heart model parameters uniformly on a 2D isotropic sheet then simulated rotors, but matching the simulated rotors to the patient's rotors remains challenging \cite{Lombardo}.

\section{Conclusion}
In this paper, we demonstrated that our computational heart model is spatially high resolution, temporally synchronous across the whole atrium, and has accurate activation patterns matching those individual patients. For high resolution, the average simulation geometry voxel is a cube of 2.47 mm length. For synchrony, the model takes in about 1,500 electrogram recordings from each patient, optimally fits parameters to the individual's atrium geometry and then generates whole-atrium synchronous activation patterns. For accuracy, the average local activation time error is 5.47 ms for sinus rhythm, 10.97 ms for flutter and tachycardia; and the average correlation is 0.95 for sinus rhythm, 0.81 for flutter and tachycardia. This is a promising result showing that the model has a potential to capture the more complex rhythms such as \ac{AF}. Such a patient-specific computational heart model allows for more efficient and effective ablation therapy to terminate AF.


\onecolumn
\appendices
\pagebreak
\section{Patient v.s. Simulation LAT Map}
\begin{table}[htbp]
\begin{center}
\begin{tabular}{ c c c } 
Patient 1 (SR) & Patient 2 (SR) & Patient 3 (SR) \\
\includegraphics[height=2.4cm]{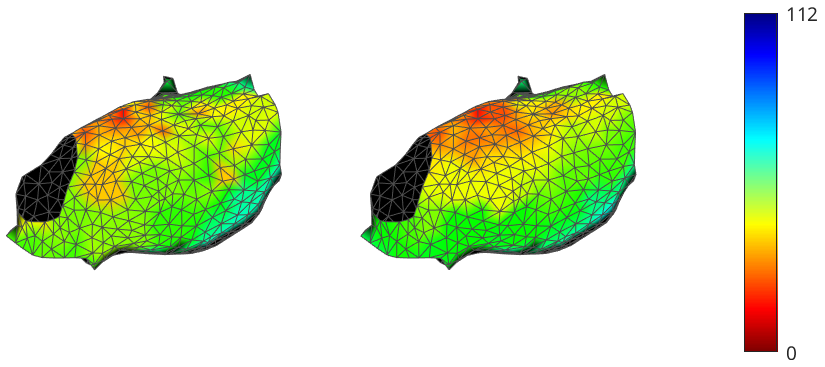} & \includegraphics[height=2.4cm]{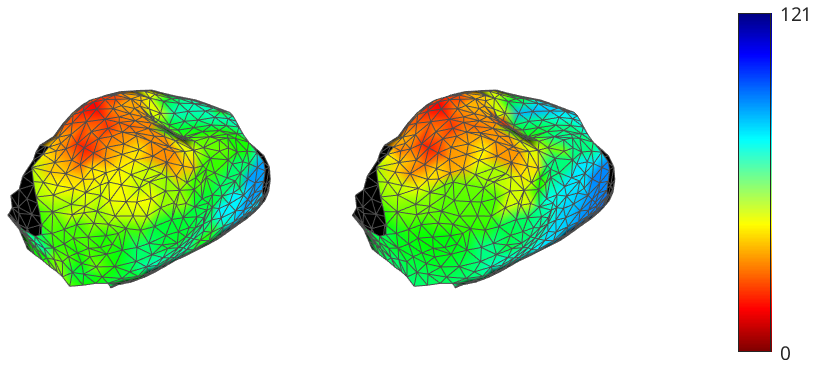} & \includegraphics[height=2.4cm]{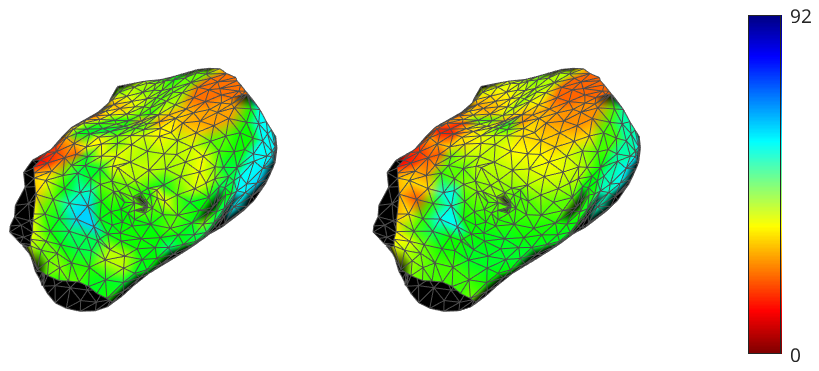} \\ 
\\ \\
Patient 4 (SR) & Patient 5 (SR) & Patient 6 (SR) \\
\includegraphics[height=2.4cm]{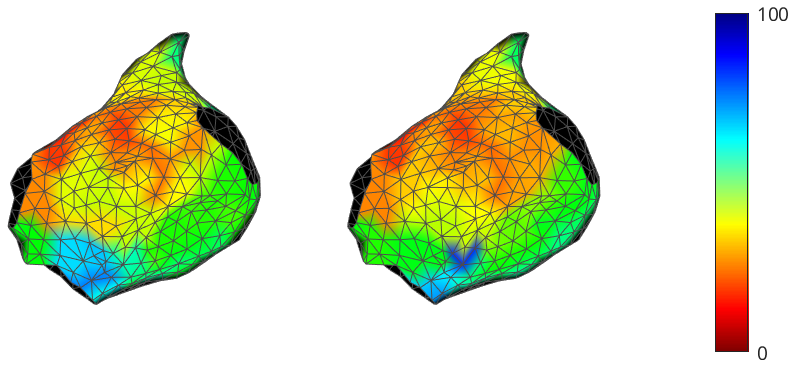} & \includegraphics[height=2.4cm]{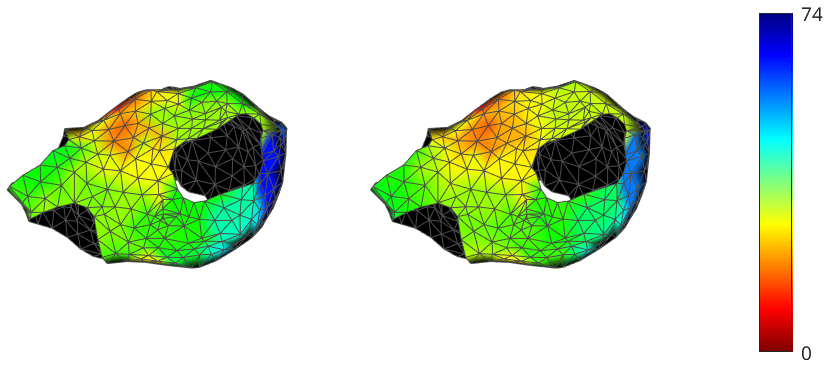} & \includegraphics[height=2.4cm]{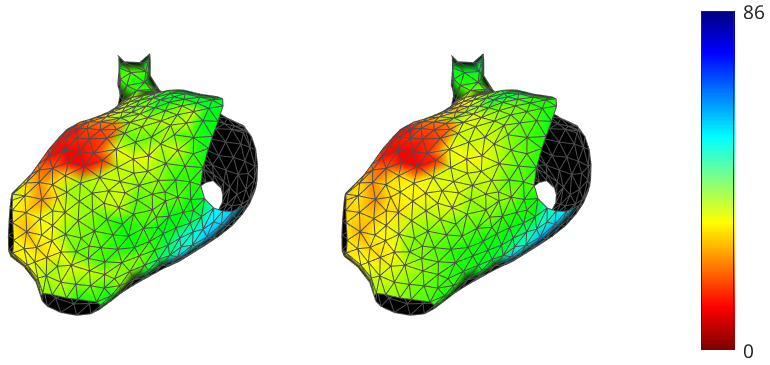} \\ 
\\ \\
Patient 7 (SR) & Patient 8 (SR) & Patient 9 (FL) \\
\includegraphics[height=2.4cm]{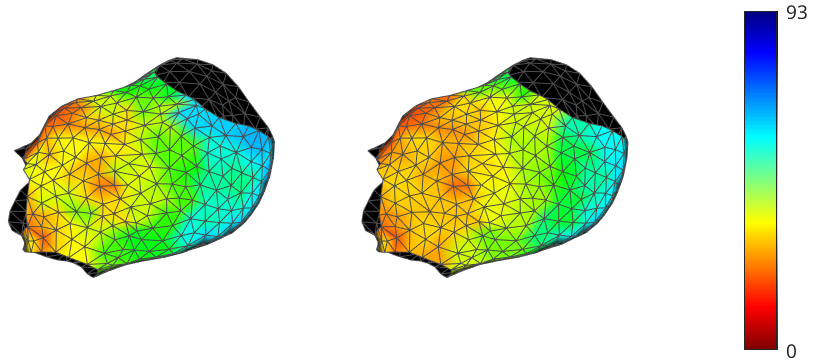} & \includegraphics[height=2.4cm]{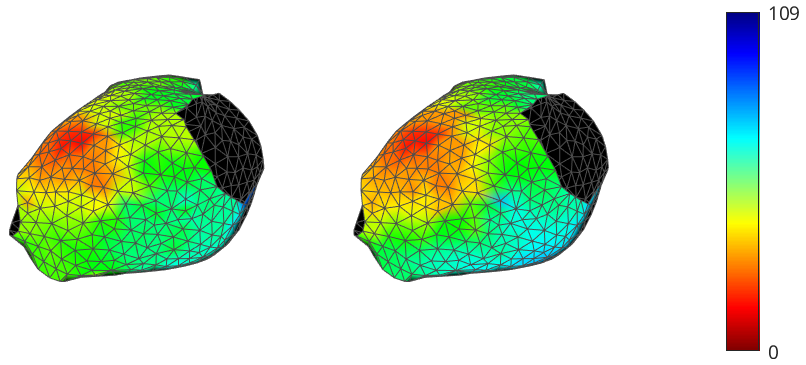} & \includegraphics[height=2.4cm]{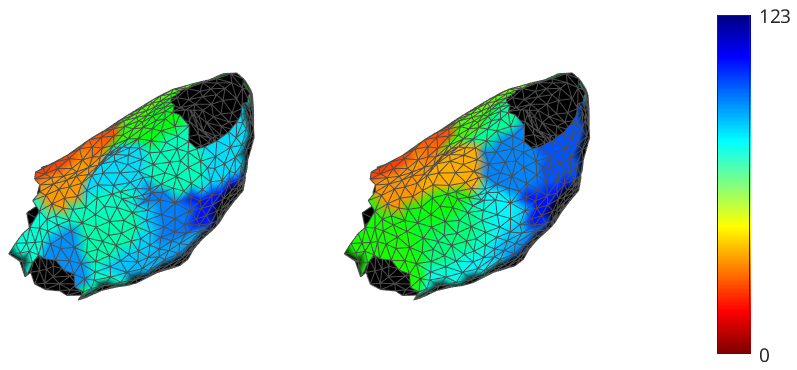} \\ 
\\ \\
Patient 10 (FL) & Patient 11 (AT) & Patient 12 (FL) \\
\includegraphics[height=2.4cm]{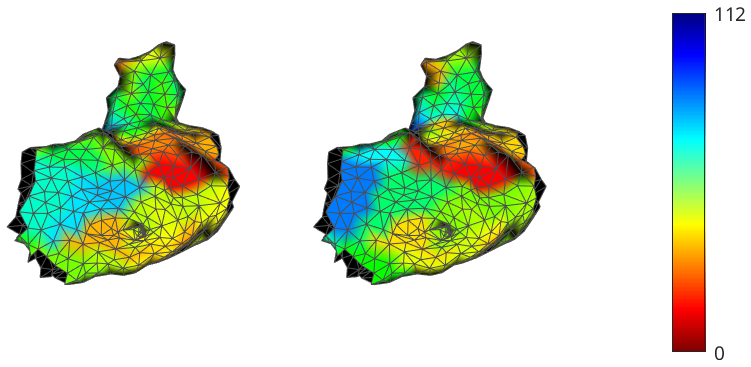} & \includegraphics[height=2.4cm]{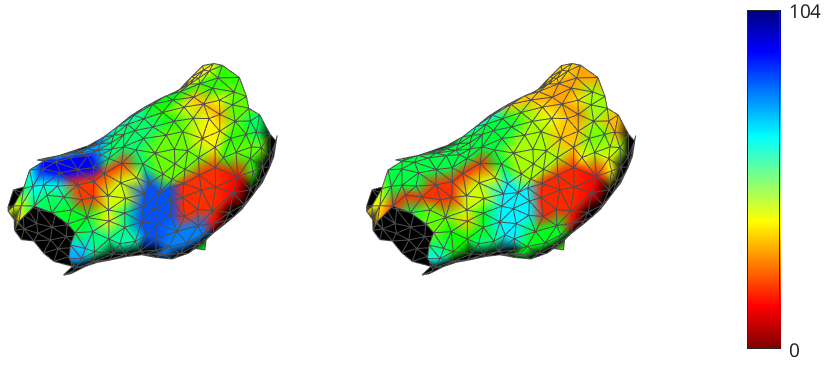} & \includegraphics[height=2.4cm]{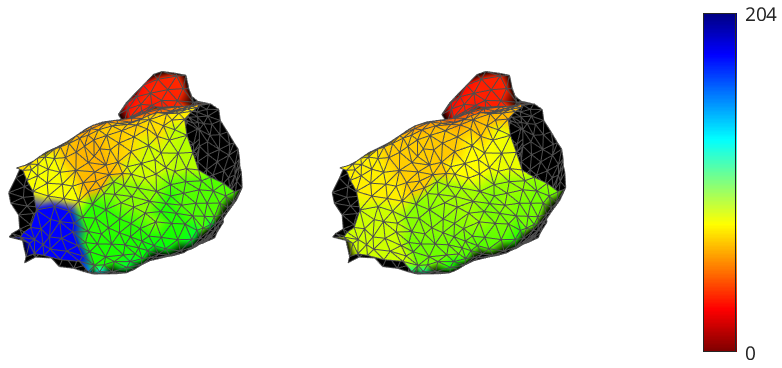} \\ 
\\ \\
Patient 13 (FL) & Patient 14 (FL) & Patient 15 (FL) \\
\includegraphics[height=2.4cm]{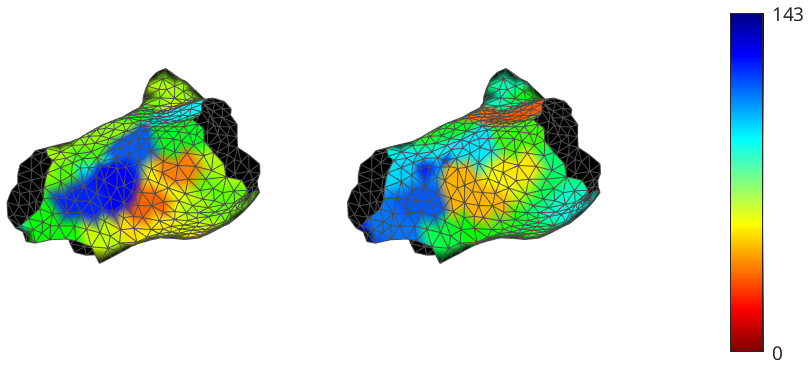} & \includegraphics[height=2.4cm]{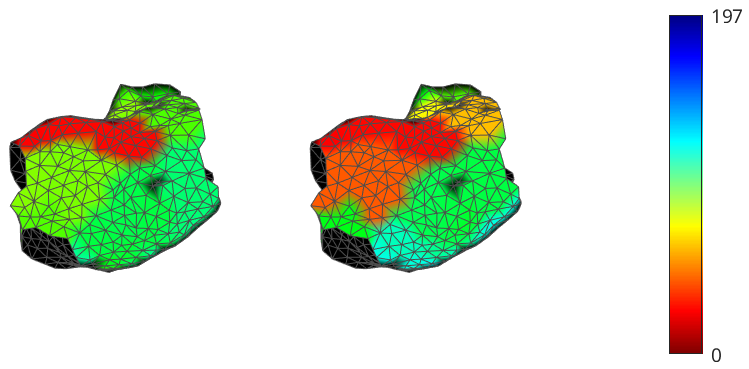} & \includegraphics[height=2.4cm]{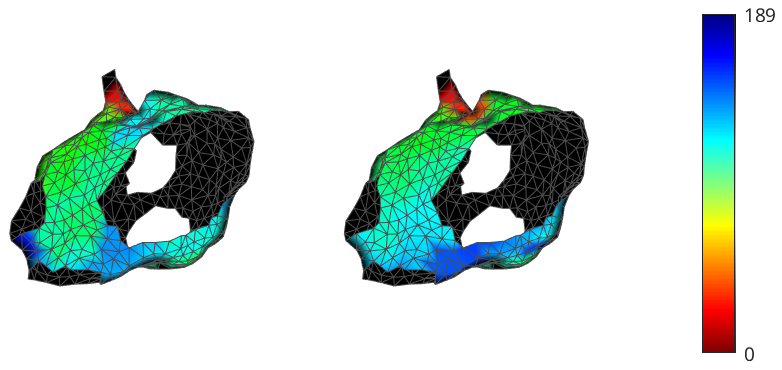} \\ 
\end{tabular}
\label{tb:performance_summary_lat_map}
\end{center}
\begin{flushleft}
Patient \ac{LAT} map is on the left, simulation \ac{LAT} map is on the right. Red represent early activation, blue represent late activation. The color bar is in unit ms. SR: sinus rhythm. FL: atrial flutter. AT: atrial tachycardia. The LAT fittings of SR maps are better than FL and AT maps. One of the reasons is that our heart model used focal sources to match rotor sources, but these two types of sources behave differently, as shown in Patient 11.
\end{flushleft}
\end{table}

\pagebreak
\section{Patient v.s. Simulation LAT Plot}
\begin{table}[htbp]
\begin{center}
\begin{tabular}{ c c c } 
Patient 1 (SR), R = 0.94 & Patient 2 (SR), R = 0.96 & Patient 3 (SR), R = 0.92 \\
\includegraphics[height=3.2cm]{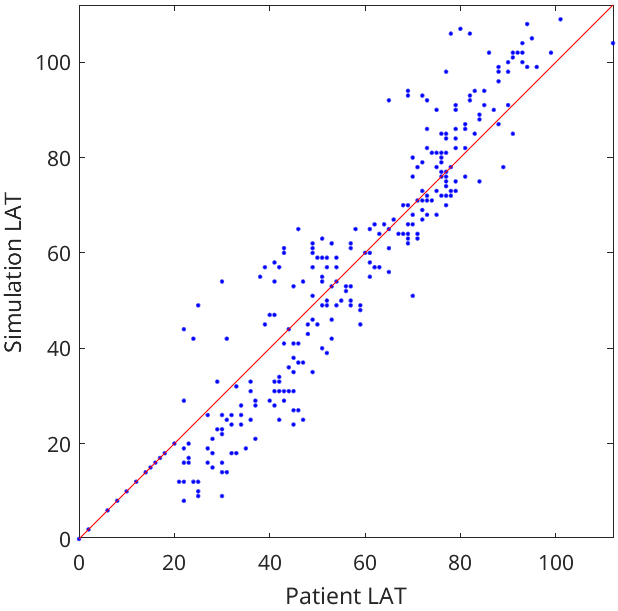} & \includegraphics[height=3.2cm]{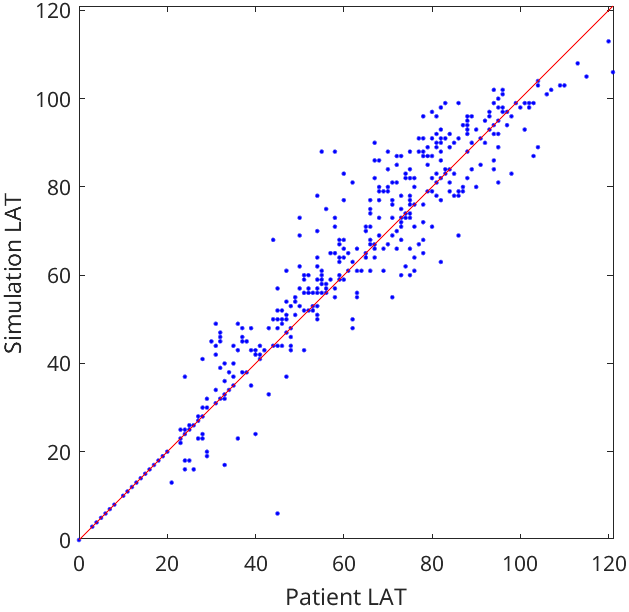} & \includegraphics[height=3.2cm]{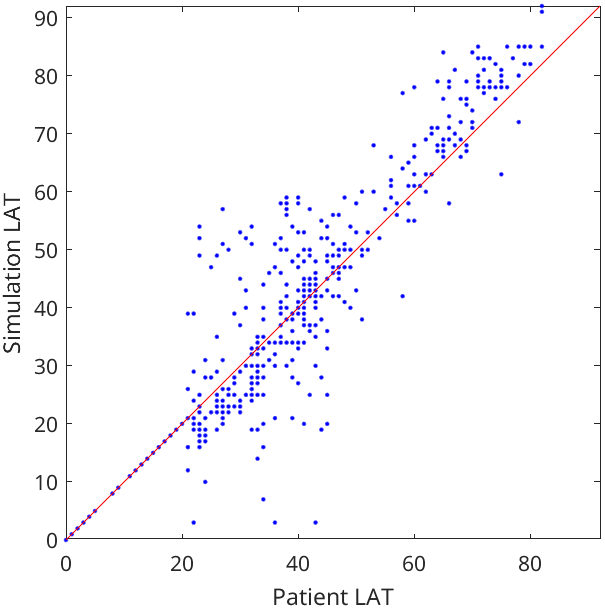} \\ 

Patient 4 (SR), R = 0.95 & Patient 5 (SR), R = 0.94 & Patient 6 (SR), R = 0.95 \\
\includegraphics[height=3.2cm]{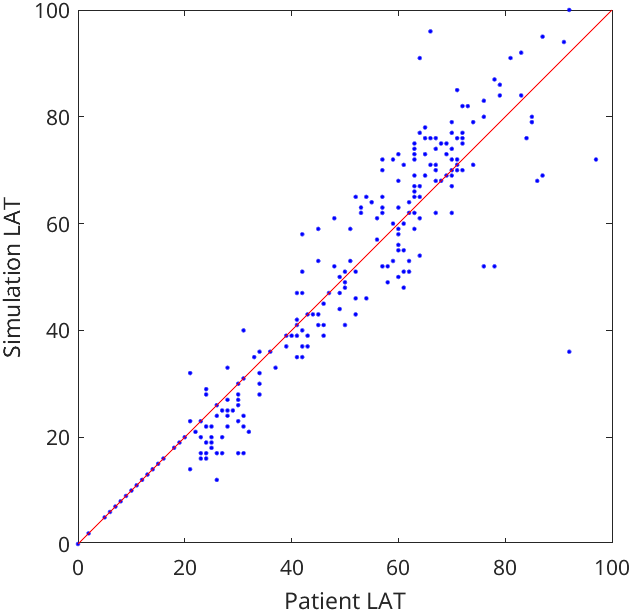} & \includegraphics[height=3.2cm]{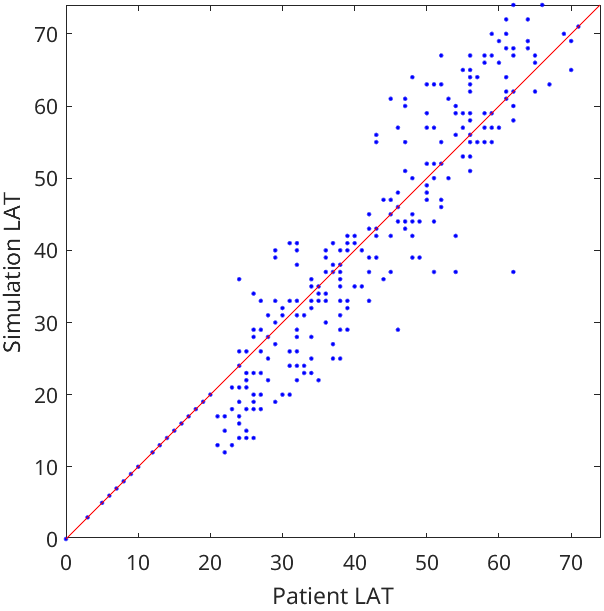} & \includegraphics[height=3.2cm]{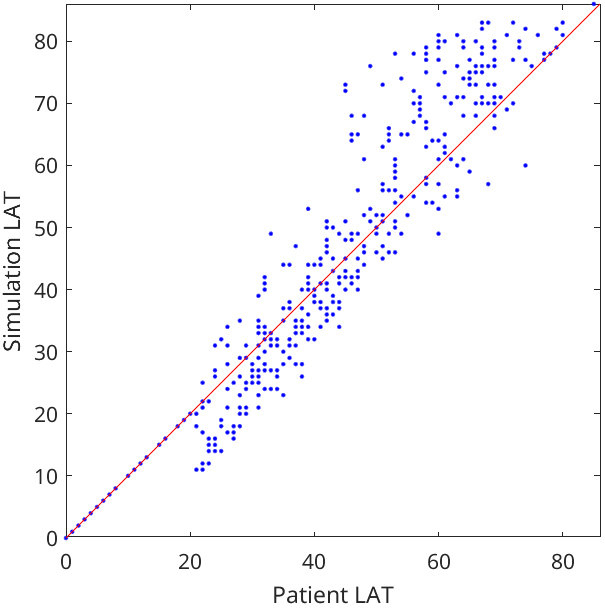} \\ 

Patient 7 (SR), R = 0.96 & Patient 8 (SR), R = 0.96 & Patient 9 (FL), R = 0.91 \\
\includegraphics[height=3.2cm]{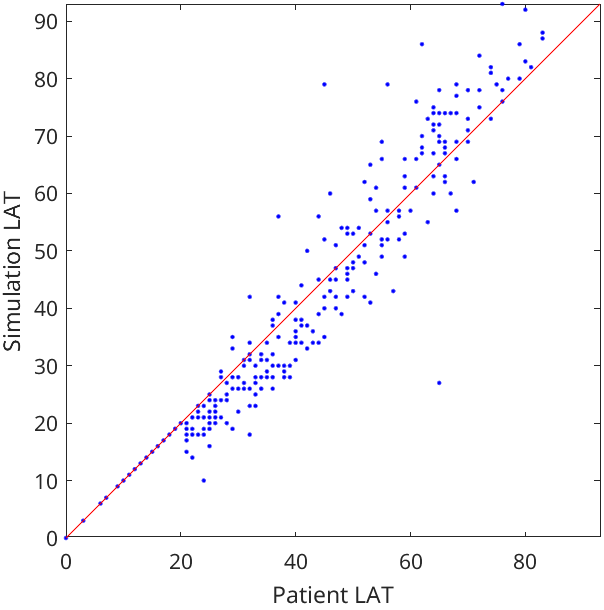} & \includegraphics[height=3.2cm]{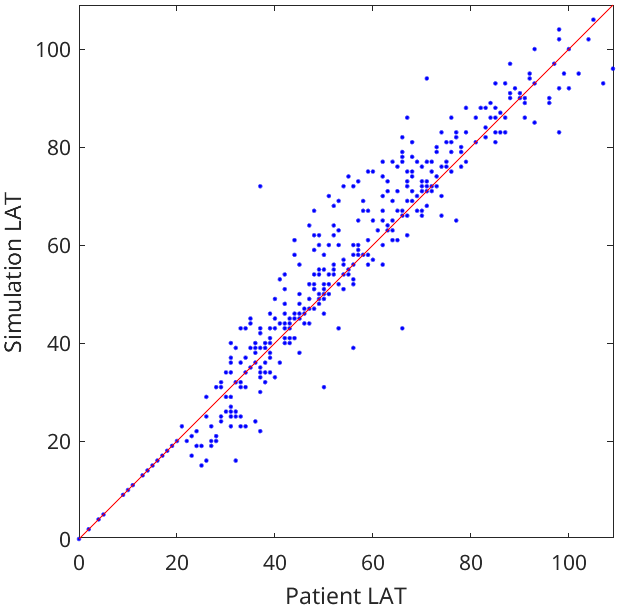} & \includegraphics[height=3.2cm]{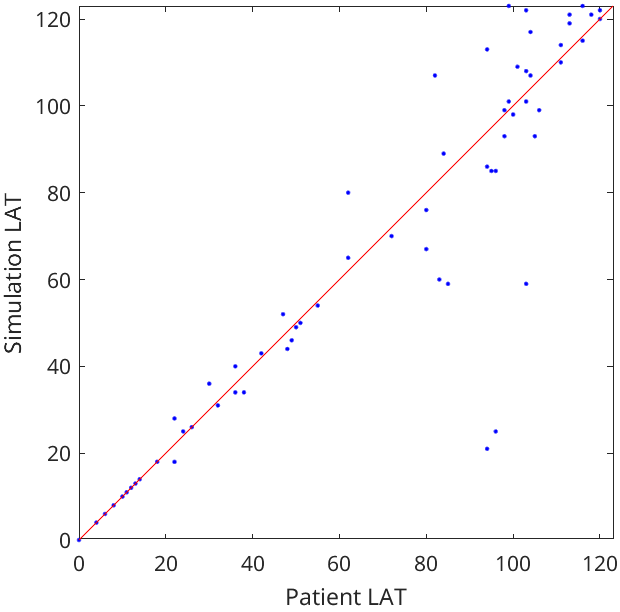} \\ 

Patient 10 (FL), R = 0.83 & Patient 11 (AT), R = 0.85 & Patient 12 (FL), R = 0.87 \\
\includegraphics[height=3.2cm]{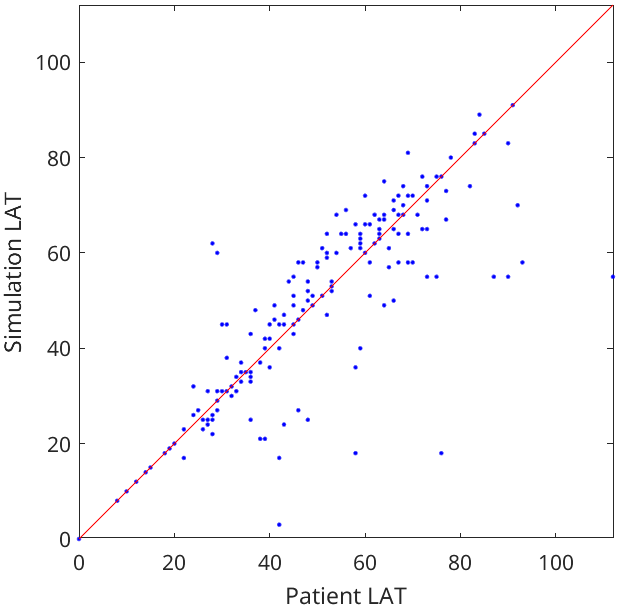} & \includegraphics[height=3.2cm]{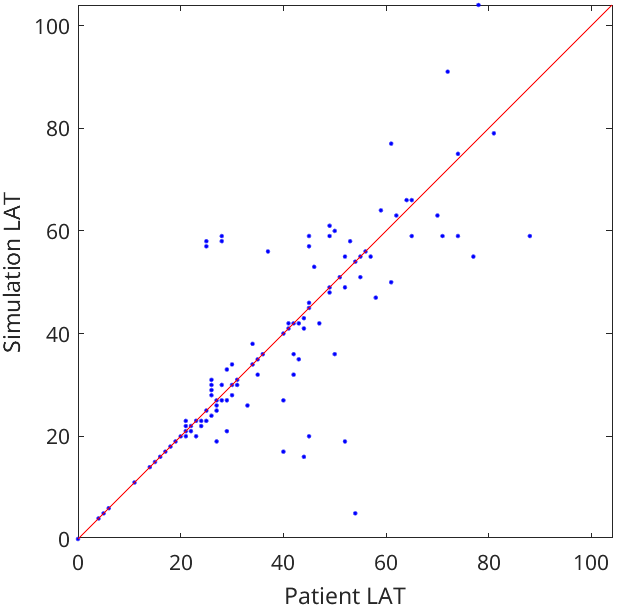} & \includegraphics[height=3.2cm]{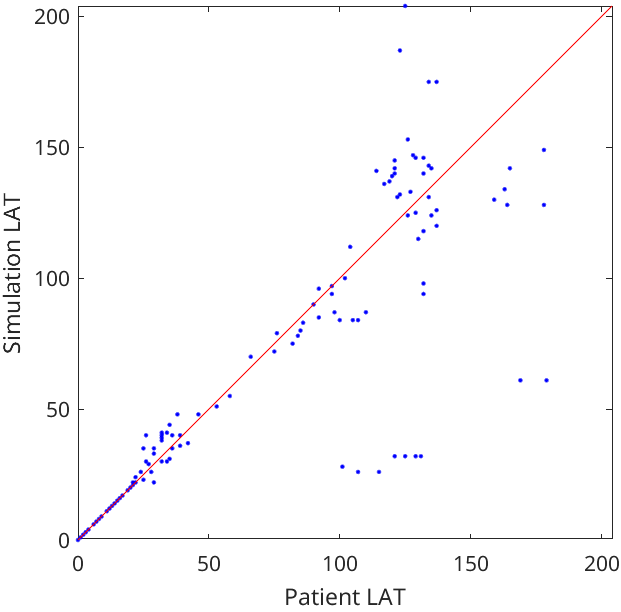} \\ 

Patient 13 (FL), R = 0.50 & Patient 14 (FL), R = 0.79 & Patient 15 (FL), R = 0.91 \\
\includegraphics[height=3.2cm]{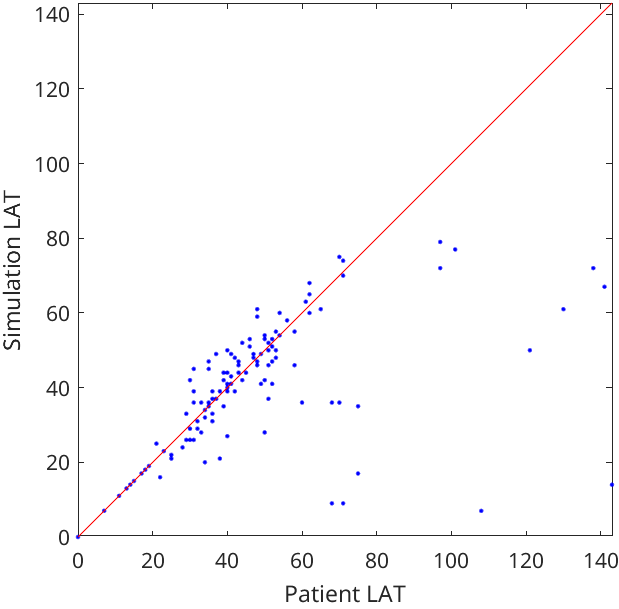} & \includegraphics[height=3.2cm]{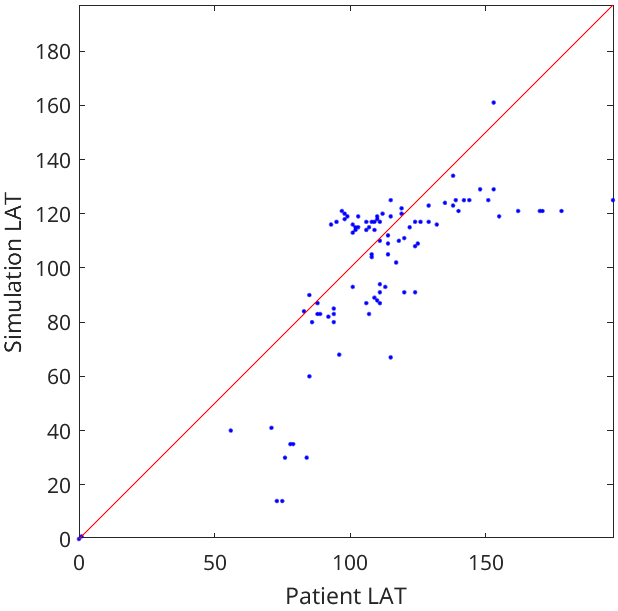} & \includegraphics[height=3.2cm]{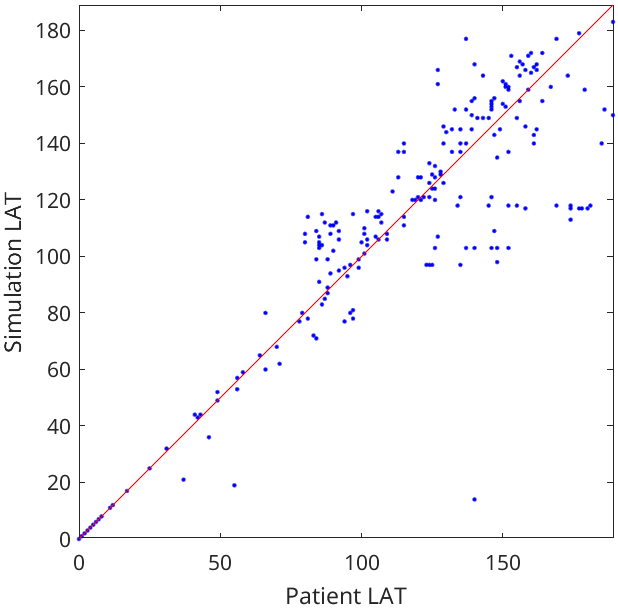} \\ 

\end{tabular}
\label{tb:performance_summary_lat_plot}
\end{center}
\begin{flushleft}
$X$ axis is patient $Y$ axis is simulated \ac{LAT}. The red line is $Y = X$.  SR: sinus rhythm. FL: flutter. AT: atrial tachycardia. R: correlation. One reason why there are less points in FL and AT plots is they have more fractionated electrograms that were excluded. One reason why FL and AT maps have lower correlation values is our heart model used focal sources to match rotor sources, but these two sources behave differently.
\end{flushleft}
\end{table}

\twocolumn

\bibliographystyle{ieeetr}
\bibliography{reference}

\end{document}